\documentclass[12pt]{iopart}
\pdfoutput=1
\usepackage{iopams}
\expandafter\let\csname equation*\endcsname\relax
\expandafter\let\csname endequation*\endcsname\relax
\usepackage{amsmath,mathtools}

\usepackage{graphicx}
\usepackage[usenames,dvipsnames]{color}

\newcommand{\ie}{{\it i.e}}
\newcommand{\eg}{{\it e.g}}
\newcommand{\etc}{{\it etc}}

\newcommand{\del}{\partial}
\newcommand{\Eqref}[1]{Eq.\ \eqref{#1}}
\newcommand{\dd}{ {\mathrm d} }

\renewcommand{\vec}[1]{\boldsymbol{#1}}
\newcommand{\mean}[1]{{\left< #1 \right>}}
\newcommand{\genL}{{\mathbb L}}
\newcommand{\Prb}{\mathrm{Pr}} 
\newcommand{\Act}{\mathcal{A}} 
\newcommand{\Frn}{\Act^{+}} 
\newcommand{\Ent}{\Act^{-}} 
\newcommand{\rf}{R} 
\newcommand{\sus}{\chi} 

\newcommand{\for}{f} 
\newcommand{\dfor}{\delta \mkern-1.5mu \for}

\newcommand{\mob}{M}
\newcommand{\Mob}{\mathbf{M}}
\newcommand{\noi}{\xi} 

\newcommand{\Dif}{\mathbf{D}}
\newcommand{\dDif}{\delta \Dif}
\newcommand{\dif}{D}
\newcommand{\ddif}{\delta \mkern-1.5mu \dif}

\newcommand{\ddi}{\delta \mkern-1.5mu \dif^{-1}}
\newcommand{\tstep}{\Delta t}

\newcommand{\caD}{{\mathcal D}}

\newcommand{\caN}{{\mathcal N}}
\newcommand{\caO}{{\mathcal O}}

\begin{document}

\title[Linear response of hydrodynamically-coupled particles]
{Linear response of hydrodynamically-coupled particles under a nonequilibrium reservoir}

\author{Cem Yolcu$^{1}$}
\author{Marco Baiesi$^{1,2}$}

\address{$^1$ Dipartimento di Fisica e Astronomia ``Galileo Galilei'',
  Universit\`a di Padova, Via Marzolo 8, 35131, Padova, Italy
} 
\address{$^2$ INFN, Sezione di Padova, Via Marzolo 8, 35131, Padova,
Italy} 

\begin{abstract}

A recent experiment driving colloids electromagnetically, by B\'erut
et al.\ \cite{ber14}, is an ideal paradigm for illustrating a linear
response theory for nonequilibrium overdamped systems including
hydrodynamic interactions and, unusually, a reservoir itself out of
equilibrium.  Indeed, in this setup one finds a nonequilibrium
environment in which the mobility and diffusivity of free particles
are not simply proportional to each other. We derive both the response
to a mechanical forcing and to temperature variations in terms of
correlations between an observable and a path-weight action. The
time-antisymmetric component of the latter turns out not to be simply
proportional to the heat flowing into the environment. These results
are visualized with simulations resembling conditions and protocols
easily realizable in the experiment, thereby tracing a path for
experimental verifications of the theory.

\end{abstract}


\submitto{JSTAT}


\maketitle

\section{Introduction}

Objects immersed in a fluid feel the presence of each other from a
distance via mediation of the fluid itself. This hydrodynamic
interaction is usually much faster than the rearrangements of the
objects so that it appears to be a long-range instantaneous effect, as
those usually modeled with energy potentials \cite{doi1988theory}.

There are hydrodynamic systems that are becoming a new paradigm of
nonequilibrium, such as bacterial motion and driven colloids
\cite{ram10}. In colonies of bacteria one is observing an instance of
active matter, where the origin of nonconservative forces is
chemically-driven self-propulsion of the single ``particles'', and can
give rise to complex phase behaviour \cite{bia12, but13}. On the other
hand, colloidal systems constitute more classical examples of systems
brought out of equilibrium by external forces.  In particular, in the
past decade, several experiments of driven colloids were performed
with the aim of realizing minimal systems in which new laws of
nonequilibrium could be tested \cite{wan02, tre04, bli06, bli07a,
  bli07}. With rotating lasers, a single particle was maintained out
of equilibrium by confining it in a toroidal region where an
additional rotational drift was imposed \cite{bli07, lut06,
  gom09}. When multiple particles were included in the toroidal
region, hydrodynamic effects led to forms of synchronization
\cite{kot10}. Another minimal system was realized recently with two
colloids trapped not far from each other, by driving the first
particle with a rapidly varying random electromagnetic field
\cite{ber14}, which thus played the role of an additional component to
the heat bath. The temperature of this particle effectively increased
by this driving and the influence was transmitted to the second
particle thanks to hydrodynamic interaction.  We will use this
experiment as our guideline to introduce a theory for the linear
response of overdamped nonequilibrium systems, with hydrodynamic
coupling and subject to a heat bath that includes a nonequilibrium
component.

Knowing how a system responds to small perturbations is at the basis
of linear response.  In equilibrium there is a celebrated result, the
fluctuation-dissipation theorem \cite{kub66}, which states that
measuring the correlations between an observable and a specific
potential in the unperturbed equilibrium regime is sufficient for
predicting how the observable would react to the addition of a tiny
amount of such energetic potential.  In the same fashion, responses to
a change in the bath temperature of a system in thermodynamic
equilibrium may be related to unperturbed correlations \cite{kub66}.
Both scenarios are actually expressions of a general relation, known
as Kubo formula. It has thus been stressed (e.g., \cite{bai09,bai13})
that to predict the response in equilibrium, the only relevant
quantity is the entropy that the perturbing entity would produce in
the dissipation toward a new equilibrium.

There are several approaches trying to generalize the
fluctuation-dissipation theory to nonequilibrium conditions, see
references in \cite{mar08, bai13}.  In some cases the knowledge of the
microscopic density of states is required \cite{aga72, fal90,
  sei10}. In other approaches there emerges the need to understand not
only entropy, which is antisymmetric with respect to time-reversal,
but also other time-symmetric dynamical aspects
\cite{cug94,lip05,bai09}.  The latter is not related to dissipation
but rather to the ``activation'' of the dynamics by means of the
perturbation. In fact, a perturbation may not only lead to an
increased or decreased tendency to dissipate but additionally to a
modified ``frenesy'' \cite{bai09,bai09b,bai10,bai13,bas15} of the
system. Such frenesy quantifies how a perturbation changes the {\em
  mean propensity} of the system to escape from a neighborhood in
state space.  For instance, in the case of discrete systems, the
frenesy indicates how the mean tendency to jump from state to state is
modified by a perturbation.  Specifically, in a given trajectory it is
not proportional to the actual number of jumps, that is, to the actual
dynamical activity~\cite{lec05,mer05,gar07}, but it is rather an
estimator of the escape rate (a kind of ``wished'' activity)
integrated in time.

Within the scope of nonequilibrium statistical physics, one usually
considers a system driven out of equilibrium by a nonconservative
force, or one where degrees of freedom interact with baths under
different equilibrium conditions. Multiple heat baths as in the latter
situation are generally employed in the study of heat transfer between
equilibrium baths at different temperatures, mediated by
mechanically-interacting degrees of freedom under their
influence~\cite{lep03,dha08}. In such models, each degree of freedom
is eventually thermalized by a \emph{separate} equilibrium
reservoir. The experimental situation briefly mentioned above
\cite{ber14} also exhibits a nonequilibrium maintained by a thermal
reservoir with different effects on different degrees of freedom, this
time introduced by the additional noise.  However, the hydrodynamic
coupling present puts a twist on the scenario. Now, the particles
receive (and dissipate) energy from (and to) the reservoir via one
another as well as directly. Therefore the action of the additional
thermal component cannot be decoupled from the fluid's, and the
exchanges with the reservoir cannot be placed in the framework of
equilibrium thermodynamics even from the point of view of the
reservoir. However, this simple system serves as a model where the
nonequilibrium, or loss of detailed balance, in the reservoir is
quantifiable, as discussed in the following section.

In the present article, we develop the linear response theory for a
system of hydrodynamically coupled degrees of freedom where thermal
equilibrium in the reservoir is broken, in the sense described in the
previous paragraph, by the presence of additional components. From the
point of view of Langevin dynamics, this amounts to a diffusivity that
is \emph{not} proportional to the mobility, as local detailed balance
would have required. We discuss first the case of perturbing the
mechanical forces, generalizing, in the necessary fashion, a
fluctuation-response relation based on path
probabilities~\cite{bai09,bai09b}. In the colloidal model system we
consider, manipulation of the artificial temperature is also an easy
and natural operation. Therefore, we deal also with fluctuation
response relations for temperature variations, further generalizing a
recent analysis of temperature response~\cite{bai14}. In both of these
cases, be it a perturbation of the deterministic influences or the
thermal ones, we see signatures of the particular out of equilibrium
condition and especially the nonvanishing correlation between the
degrees of freedom due to hydrodynamic coupling. We illustrate our
findings by data from numerical experiments.

Furthermore, from considering path probabilities, the nonequilibrium
situation maintained by the non-proportionality between diffusivity
and mobility hint at possible caveats on identifying the
time-antisymmetric part of the path action with dissipated
heat. Indeed, the relationship between dissipated heat and entropy is
a temperature, which is not possible to identify unequivocally by
simply looking at the reservoir, since the reservoir is not made up of
separate heat baths at well-defined temperatures. Within the distinct
framework of the experimental paradigm of Ref.~\cite{ber14}, it is
possible to write an explicit expression for the dissipation, upon
which we see that it contains terms that are not of the form of heat
in the sense of stochastic thermodynamics, but those that couple the
forces and displacements across different particles. This seems to be
a form of housekeeping entropy \cite{oon98, hat01,spe06}, incurred by
the breaking of detailed balance by the artificial temperature on one
particle.

The article is organized as follows: Sec.\ \ref{sec:framework}
establishes the Langevin dynamics framework within which we address
the problem. Here, we show explicitly how the artificial noise is
encoded in the diffusivity matrix, and proceed with a concise review
of the experimental paradigm along with its placement into the
framework. Sec.\ \ref{sec:path} begins with a review of the path
integral framework in the context of linear response, and moves on to
explicit expressions for the particular physical situation under
consideration. From this perspective, the aforementioned peculiarities
of heat and entropy production are discussed. In
Sec.\ \ref{sec:resp.f}, we derive the linear response formula for
deterministic perturbations and apply it to a situation where optical
trap strengths are manipulated. The section ends with results and
discussion of numerical experiments measuring the associated
susceptibility of the energies stored in the traps. Similarly, in
Sec.\ \ref{sec:resp.T} we discuss the linear response to variations of
the artificial temperature and provide results of illustrative
numerical experiments, before we conclude the article.

\section{Stochastic framework} \label{sec:framework}

Let us begin with a concise discussion on the stochastic dynamics of the
type of systems investigated in this article, and establish aspects
of notation. First we explain how hydrodynamic coupling and
the presence of an additional (electromagnetic) random forcing should
be incorporated. Afterwards, we specialize these arguments to the
particular experimental situation that serves as the central physical
paradigm for the scope of this article.

\subsection{Langevin dynamics}

The degrees of freedom describing the system are the coordinates 
$x_i(t)$ of micrometer-sized particles in an aqueous solvent. At this
scale, inertia becomes irrelevant; momenta are excluded and the
dynamics is overdamped. Therefore, the dynamics will be determined by
a system of coupled overdamped Langevin equations, such as
\begin{align}
  \dot{x}_i = \mob_{ij} \for_j (\vec x ) + 
  (\sqrt{2 \dif})_{ij} \noi_j \label{eq:Lang.ind} \ ,
\end{align}
where time dependences of $x_i (t)$ and $\noi_i (t)$ have been omitted
for brevity of notation, and Einstein's summation convention is
understood.\footnote{On occasion, we will use matrix notation instead
  of indices. In these instances we will use boldface symbols in place
  of collections of numbers, be it vectors (as in the argument of
  $\for (\vec x)$) or matrices.} The vector $\noi_i (t)$ of colorless
Gaussian differential noises has uncorrelated components, \ie.,
$\left< \noi_i (t) \noi_j (t') \right> = \delta_{ij}
\delta(t-t')$.
Any correlations in the random forcing, therefore, are captured in the
(force-free) diffusion matrix $\dif_{ij}$, which is
symmetric.\footnote{The expression $\sqrt{\dif}$ stands for a matrix
  which satisfies $(\sqrt{\dif})_{ik} (\sqrt{\dif})_{jk} =
  \dif_{ij}$.
  As such, the matrix $\sqrt{\dif}_{ij}$ is not unique; there is
  freedom to choose $\sqrt{\dif}_{ij} \to \sqrt{\dif}_{ik} O_{kj}$
  with $O_{ij}$ an orthogonal transformation, without affecting the
  physical content \cite{gardiner}.} The symmetric (free-particle) 
mobility matrix
$\mob_{ij}$ embodies inverse drag coefficients originating from the
dissipative nature of the fluid, and $\for_i (\vec x)$ is the
deterministic force on the degree of freedom $i$. The mobility and
diffusion coefficients are assumed to be state-independent. This is an
approximation that is suitable for the particular physical paradigm we
focus on, as explained later in Sec.\ \ref{sec:expara}.

\subsection{A multicomponent bath with hydrodynamic coupling}

Hydrodynamic coupling implies that forces on one degree of freedom
affect the motion of other degrees of freedom via propagation through
the fluid. Hence, the mobility matrix $\mob_{ij}$ is not
diagonal. Since the same applies to the random forces exerted by the
fluid, the random forcing term also embodies correlation between
different degrees of freedom, encoded in the off-diagonal elements of
the diffusivity matrix $\dif_{ij}$.

In this article, we deal with an aqueous environment at a temperature
$T$. The dissipative and fluctuating effects of the fluid, namely the
damping of motion (encoded in ${\mob^{-1}}_{ij}$) and the stochastic
energy pumping (encoded in $\dif_{ij}$), balance each other. This is
formalized by the so-called second fluctuation-dissipation 
theorem \cite{kub66,mar08}
(with Boltzmann's constant set to $k_{\rm B}=1$) as,
\begin{equation}
  \dif^{\rm aq}_{ij} = T \mob_{ij} \ . \label{eq:2ndFT}
\end{equation}
However, the aqueous environment is considered as just one component
of a larger, ``multicomponent'' bath. This is inspired by recent
experiments \cite{ber14} where colloids were placed in solution via
optical tweezers subjected to random oscillations, effectively
thermalizing the particles at different temperatures. A conceptual
description where the bath consists of aqueous and electromagnetic
components suits this system. In such a case with multiple
thermalizing influences, the proportionality \eqref{eq:2ndFT} between
mobility and diffusion is broken. One can separate
the diffusion matrix into separate contributions from the fluid and
the electromagnetic field.

The aforementioned random oscillation of the optical traps was
generated through white noise, and hence enters the dynamics of the
system as a stochastic force in addition to that of the aqueous
environment. In other words, we have
\begin{align}
  \dot{x}_i = \mob_{ij} \for_j (\vec x) + 
  {\sqrt{2 \dif^{\rm aq}}}_{ij} \noi^{\rm aq}_j +
  {\sqrt{2 \dif^{\rm em}}}_{ij} \noi^{\rm em}_j
  \ , \label{eq:Lang.ind.exp}
\end{align}
appropriately labeling the parts due to the aqueous environment and
the electromagnetic field of the optical traps. One can write a
completely equivalent equation,
\begin{align}
   \dot{x}_i = \mob_{ij} \for_j (\vec x) + 
   {\sqrt{2 \!\left( \dif^{\rm aq} + \dif^{\rm em} \right)}}_{ij} 
   \noi_j \ , 
\end{align}
by adding the diffusion (or covariance) matrices, as a result of the
fact that the sum of two independent Gaussian random variables is
still a Gaussian random variable whose variance is the sum of the two
original variances. Hence, the multicomponent reservoir is
described by a diffusion matrix
\begin{align}
  \dif_{ij} = \dif^{\rm aq}_{ij} + \dif^{\rm em}_{ij} \ . \label{eq:D.exp}
\end{align}
Underlying this additive separation of diffusion matrices
\eqref{eq:D.exp}, or of stochastic forces \eqref{eq:Lang.ind.exp},
there is one assumption: Despite the influence of the electromagnetic
components, the aqueous bath remains in equilibrium at temperature
$T$, still satisfying the fluctuation relation \eqref{eq:2ndFT}. This
assumption was also adhered to in the analysis of Ref.\ \cite{ber14},
where agreement with experimental measurements was demonstrated using
(cross-)correlations of the coordinates in a steady state.

The remaining question is the form of the diffusion matrix
$\dif^{\rm em}_{ij}$. In practice, it is not the components of
$\dif^{\rm em}_{ij}$ that are under direct experimental control, but
the artificial random force, $\for^{\rm em}_{i}$, that is acting on
each degree of freedom $i$. More specifically, the experimenter knows
the correlation function,
\begin{align}
  \mean{\for^{\rm em}_i (t) \for^{\rm em}_j (s)} = 2 C^{\rm em}_{ij}
  \delta (t-s) \ , \label{eq:emcorr}
\end{align}
of the colorless Gaussian random forcing. For instance, each component
might be statistically independent leading to a diagonal correlation
matrix $C^{\rm em}_{ij}$, or all but one might vanish leading to a
projector-like matrix, the latter being the specific case realized in
the experiment of Ref.\ \cite{ber14}. It is then straightforward to
relate the diffusion matrix $\dif^{\rm em}_{ij}$ to the correlation
matrix $C^{\rm em}_{ij}$ by simply noting that according to the
equation of motion \eqref{eq:Lang.ind.exp}, the random force
$\for^{\rm em}_i$ is given as
\begin{align}
  \for^{\rm em}_{i} = (\mob^{-1} \sqrt{2 \dif^{\rm em}})_{ij} 
  \noi^{\rm em}_j \ . \label{eq:force.em}
\end{align}
Requiring this expression to yield the same correlations as
\Eqref{eq:emcorr} fixes the diffusion matrix to be
\begin{align}
  \dif^{\rm em}_{ij} = (\mob C^{\rm em} \mob)_{ij} \ .
\end{align}

Thus, we have described a system where degrees of freedom are subject
to overdamped dynamics in a fluid medium in thermal equilibrium,
including hydrodynamically mediated correlations, further under the
influence of stochastic forces originating from a source separate from
the fluid medium. In this fashion, the resulting system is one in
contact with a reservoir out of equilibrium due to multiple
components. The equation of motion \eqref{eq:Lang.ind} applies to such
a system when the diffusion matrix is decomposed as in
\Eqref{eq:D.exp}, which reads
\begin{align}
  \dif_{ij} = T \mob_{ij} +
  (\mob C^{\rm em} \mob)_{ij} \ , \label{eq:D.exp2}
\end{align}
when expressed explicitly in terms of the properties of the fluid and
those of the external noise. We note that the arguments above
justifying this decomposition do not rely on the coefficients being
position-independent, even though we will not be considering that
general case.

\subsection{Experimental paradigm} \label{sec:expara}

A recent experiment \cite{ber14} is suitably described in the
framework laid down above and is chosen to be our central paradigm. A
pair of colloids (radius $R \mathbin{\approx} 1\,\mu$m) were held in
narrow optical traps a few microns apart inside water, such that
hydrodynamic coupling would not be negligible. The effective stiffness
constants ($\kappa$) of the optical traps were $\kappa
\mathbin{\approx} 4\,$pN/$\mu$m, placing the average deviation,
$\sqrt{\mean{x^2}}$, of each coordinate roughly at $\sqrt{T/\kappa}
\mathbin{\approx} 3 \times 10^{-2} \mu$m, under a typical room
temperature $T\mathbin{=}4$\,pNnm ($k_{\rm B} =1$).  Denoting the
inter-trap distance by $d$, the smallness of $\sqrt{\mean{x^2}}/ d$
allows an approximation to lowest order in this ratio, leading to two
simplifications: (i) The two directions orthogonal to the line joining
the traps can be neglected, which lessens the algebraic burden. (ii)
The dependence on coordinates of the mobility and diffusion matrices
can be ignored, using their values corresponding to the state when
each particle is situated at the minimum of their trap ($x_i
\mathbin{=} 0$). That is, $\mob_{ij} (\vec x) \mathbin{=} \mob_{ij} (
\vec 0) \mathbin{=} \mob_{ij}$, and $\dif_{ij} (\vec x) \mathbin{=}
\dif_{ij} (\vec 0) \mathbin{=} \dif_{ij}$.  For the development of the
linear response to temperature variations, for the moment it seems
particularly challenging to deal with a state-dependent noise
prefactor, making this latter simplification useful. This is also in
line with the approach taken by the authors of Ref.\ \cite{ber14} in
their theoretical analysis of the experiment.

Under this approximation, there remain only two degrees of freedom,
$\vec x \mathbin{=} (x_1, x_2)$, and the forces on the particles are
described by the (potential) energy function
\begin{align}
  U (\vec x) &= U_1(x_1) + U_2(x_2) \nonumber\\
             &= \frac 12 \kappa_1 x_1^2 + \frac 12 \kappa_2 x_2^2 \ .
\end{align}
The energy is divided into two terms, each depending on either $x_1$
or $x_2$, hence the forces $\for_i \mathbin{=} -\del_i U (\vec x)$ are
transmitted between the degrees of freedom only by the fluid, which is
captured by the mobility matrix. The mobility matrix corresponding to
hydrodynamically coupled identical spherical colloids was discussed,
for example, in Refs.\ \cite{tok94, tok95}. Applied to the case at
hand, with two particles moving along the same spatial dimension, we
have
\begin{align}
  \Mob = \frac 1\gamma \begin{bmatrix} 1 & \epsilon \\
    \epsilon & 1 \end{bmatrix} \ , \label{eq:mob}
\end{align}
where $\gamma$ is the drag coefficient for a spherical colloid of
radius $R$, and $\epsilon$ is given as \cite{tok94, tok95}
\begin{align}
  \epsilon = \frac{3 R}{2 d} - \frac {R^3}{d^3} \ 
  + \ldots ,
\end{align}
with $d$ the distance between the two colloids. These are the first
two terms in a multipole expansion in the ratio $R/d$, sometimes
referred to respectively as Oseen, and Rotne-Prager terms
\cite{rot69}. However, for our purposes, the distinction will not be
important.

The crucial ingredient of the experiment of Ref.\ \cite{ber14} is the
presence of the external stochastic force affecting one
coordinate---say coordinate 1---realized through the application of a
white noise to the circuitry responsible for positioning the first
optical trap. In the framework laid down in the previous section, this
amounts to an electromagnetic forcing with a correlation matrix
\eqref{eq:emcorr} of the form $C^{\rm em}_{ij} \mathbin{=} 0$ unless
$i \mathbin{=} j \mathbin{=} 1$. Further identifying an effective
temperature $T_{\rm em}$ (which was called $\Delta T$ in
Ref.\ \cite{ber14}) with the power of the noise (see
Ref.\ \cite{mar13a} for further details), one has
\begin{align}
  \Dif^{\rm em} = \Mob \begin{bmatrix} \gamma T_{\rm em} &0\\0&0
  \end{bmatrix} \Mob \ . \label{eq:Dem}
\end{align}
In other words, the two thermalizing influences present in the problem
can be described by the diffusion matrix \eqref{eq:D.exp2}
\begin{align}
  \Dif = T \Mob + T_{\rm em} \Mob \begin{bmatrix} \gamma&0
    \\0&0 \end{bmatrix} \Mob \ , \label{eq:Dif.2}
\end{align}
which of course coincides with the analogous coefficient matrix
\begin{align}
  \Dif = \frac 1\gamma \begin{bmatrix}
    T + T_{\rm em} & \epsilon (T + T_{\rm em}) \\
    \epsilon (T + T_{\rm em}) & T + \epsilon^2 T_{\rm em}
    \end{bmatrix} \ , \label{eq:Dif.expl}
\end{align}
that was identified from the associated Fokker-Planck equation in
Ref.\ \cite{ber14}.

We will see in further sections of the article that the ``inverse
temperature matrix'' $(\dif^{-1} \mob)_{ij}$ will play a significant
role. For reference, let us provide its explicit expression
corresponding to this particular situation:
\begin{align}
  \Dif^{-1} \Mob = \begin{bmatrix} \frac {1} {T + T_{\rm em}}
      & - \frac{T_{\rm em}}{T} 
    \frac{\gamma \mob_{12}} {T + T_{\rm em}}
        \\ 0 & \frac{1}{T} \end{bmatrix} \ . \label{eq:itm}
\end{align}
Under equilibrium reservoir conditions ($\mathbin{T_{\rm em} = 0}$),
this is simply equal to $\delta_{ij}/T$. When there is no hydrodynamic
coupling ($\mob_{i \neq j} \mathbin{=} 0$), it is diagonal but with
unequal entries, which would correspond to the typical nonequilibrium
situation of several {\em decoupled} heat baths at unequal (inverse)
temperatures.

In the following sections we will investigate
the response of a state observable of the system to a small
perturbation of either the deterministic or the stochastic (thermal)
forces on the degrees of freedom. For instance, a deterministic
perturbation ($\mathbin{\for_i \rightarrow \for_i + \dfor_i}$) could
be realized experimentally with ease by varying the stiffness of the
optical traps ($\mathbin{\kappa_{i} \rightarrow \kappa_{i} + \delta
  \kappa_{i}}$). On the other hand, the experimental setup described
above also allows the perturbation of the system thermally in a
distinct sense: The effective temperature of external noise can in
principle be varied virtually as quickly as desired. Therefore, we
will investigate the response of the system to small variations of the
noise amplitudes encoded in $\dif_{ij}$ as a result of a variation of
the effective temperature, $T_{\rm em} \rightarrow T_{\rm em} + \delta
T_{\rm em}$, according to \Eqref{eq:Dif.2} for instance.

\section{Path functionals and fluctuation-response relations} \label{sec:path}

We begin this section by reviewing the path integral framework for
linear response, under quite general conditions first, without
referring explicitly to hydrodynamic coupling and external noise. This
reduces issues of notation and makes the article more self-contained.

\subsection{Linear response} \label{sec:linear.general}

The central quantity of interest is the first order change in the
expectation value of an observable upon a small perturbation of
parameters. Namely, we seek $\delta \!\left< \caO(t) \right>$, where
$\caO (t)$ is an observable evaluated at time $t$ and the symbol
$\delta$ for first order variation. The variation of the average
arises from the modification of the associated probability weights,
which becomes apparent when written as a path integral. Denoting
entire trajectories with the symbol $\vec x ()$,
\begin{align}
  \left< \caO (t) \right> = \int \caD \vec x () \Prb [\vec x()]
  \caO(t) \ ,
\end{align}
where $\caD \vec x() \Prb [\vec x()]$ is the probability measure in
the vicinity of $\vec x()$ in trajectory space. Reserving an explicit
expression for the path probability for a subsequent subsection, the
variation of the expectation value simply follows as 
\begin{align}
  \delta \! \left< \caO (t) \right> =& \int \caD \vec x ()
   \delta\Prb [\vec x()]  \caO(t)
   = \int \caD \vec x() \Prb [\vec x()] \caO (t) \,
   \delta\! \log \Prb [\vec x()] \ . \label{eq:response1}
\end{align}
The path weight $\Prb [\vec x()]$ is often expressed in terms of a
{\em path action}~\cite{leb99,mae99,bai09} 
defined as 
\begin{align}
  \Act [\vec x()] \equiv -\log \Prb [\vec x()]\ , \label{eq:Actdef}
\end{align}
upon which the response \eqref{eq:response1} assumes the form of a
correlation of the observable with the action {\em excess} generated
due to the perturbation:
\begin{align}
  \delta \mean{ \caO (t)} = - \left< \caO (t) \, 
  \delta \Act [\vec x()] \right> \ . \label{eq:response}
\end{align}
The average is in the \emph{unperturbed} ensemble.

In studies of linear response, it is customary to refer to a response
function (or a unit-impulse response) akin to a Green function, which
conceptually separates the response from the particular
time-dependence of the perturbation sourcing the response. When a
global scalar parameter $\lambda$ of the system is
subjected to a time-modulated perturbation $\delta \lambda (t)$, the response
function $\rf_{\caO}^{\lambda} (t,s)$ is defined through
\begin{align}
  \delta \mean{\caO (t)} = \int_{-\infty}^t \dd s\,
  \delta \lambda (s) \rf_{\caO}^{\lambda} (t,s) \ .
  \label{eq:rf.green}
\end{align}
The integration limits reflect causality. Consequently, the response
function becomes the functional derivative
\begin{align}
  \rf_{\caO}^{\lambda} (t,s) = \frac{\delta \mean{\caO (t)}} {\delta
    \lambda(s)} {=} - \mean { \caO (t)
    \frac {\delta \Act [ \vec x()]} {\delta \lambda (s)}}
  \ . \label{eq:rf.fnc}
\end{align}
As with the average, the derivatives are also evaluated at zero
perturbation $\delta \lambda=0$. With all variations truncated after
first order in a linear response theory by construction, it is to be
understood that what remains {\em after} the variation is to be
evaluated at the unperturbed point. 

A special case of practical interest is when the perturbation is
step-like, that is,
$\delta \lambda (t) = \delta \lambda \, \vartheta (t)$ with
$\vartheta (t)$ being the Heaviside step function. The integrated
response under such circumstances---see \Eqref{eq:rf.green}--- allows
one to define a {\em susceptibility},
\begin{align}
  \frac {\del \mean{\caO (t)}} {\del \lambda} = 
  \int_0^t \dd s\, \rf_{\caO}^{\lambda} (t,s) 
  \equiv \sus_{\caO}^{\lambda} (t) \ , \label{eq:sus.rf}
\end{align}
more in the spirit of quantities such as compressibility, heat
capacity, \etc. 

\subsection{Path action} \label{sec:paction}

At this point, we give an explicit expression for the path action of
our system with hydrodynamic coupling and external noise. This is done
adopting the Stratonovich convention of stochastic integrals. This
interpretation is commonly regarded as the ``native'' one for
stochastic differential equations proposed for physical scenarios
where the noise term signifies the accumulated microscopic forces
during each time step. Although in the formal limit of infinite time
resolution ($\tstep \rightarrow 0$) it is possible to shift between
different noise interpretations, in experiments one acquires
trajectory data in the form of {\em discrete} time series, with a
finite $\tstep$. We would like our expressions for the averages in
\Eqref{eq:response} to be readily applicable to discrete experimental
data, and hence follow the Stratonovich convention from the outset 
(see also the Appendix).

Based on the fact that the random variable $\tstep \, \noi_i (t)$ has
a Gaussian probability distribution of variance $\tstep$, the
probability weight $\Prb [\vec x()] = \exp \!\left( - \Act[\vec x()]
\right)$ can be written as the exponential of a path integral
quadratic in path increments.  Using the equation of motion
\eqref{eq:Lang.ind}, the action \eqref{eq:Actdef} hence
becomes
\begin{align} 
  \Act[\vec x()] = \log \caN^{\dif}_{\tstep} + \frac 14 \int \dd s
  \left[ (\dot{x}_i - \mob_{ik} \for_k) (\dif^{-1})_{ij} (\dot{x}_j -
    \mob_{jl} \for_l)+ 2 \mob_{ij} \del_i\for_j \right]
  \ , \label{eq:action}
\end{align}
where time and coordinate dependences have been kept implicit. The
last term $\sim \del_i \for_j (\vec x)$ is a consequence of the
Stratonovich interpretation of the stochastic integrals. In general,
without the approximation of state-independent coefficients, the
expression for the Stratonovich path action is much more complicated
\cite{arn00, lan82book}. Also note that in the expression above,
neither the integration limits nor an initial distribution were
included explicitly, as they need not be determined at this
point. Therefore, the results we obtain have an understood dependence
on the initial density of states, and may be used both for steady
states and for transient dynamics.

The integral above should be taken as convenient notation for the
continuum limit of a sum over discrete time steps of size $\tstep$,
rather than a standard continuum integral. With this perspective,
$\caN^{\dif}_{\tstep}$ is a normalization factor, independent of
trajectory, which is regularized by a finite $\tstep$ and depends on
the dynamics only through the diffusivity $\dif_{ij}$:
\begin{align}
  \log \caN^{\dif}_{\tstep} = \frac 12 \sum_{\tstep} \operatorname{Tr} \log
  (4 \pi \tstep \dif_{ij}/\ell^2) \ . \label{eq:norm}
\end{align}
Here, $\ell$ is an inconsequential length for correct dimensionality,
and the sum is over time slices which were left without indices for the sake
of notation. When the diffusion matrix is constant, this dangerously
singular normalization becomes irrelevant, while it needs to be taken
into account when perturbations on the temperature or diffusivity are 
considered~\cite{bai14}, as we will see later.

Given this explicit form \eqref{eq:action}, it is easy to decompose
the action into parts with odd ($-$) and even ($+$) parity under a
time-reversal transformation: Assuming forces of even parity, one need
only look at the power of $\dot{x}$ in each term. The decomposition
\begin{align}
  \Act [\vec x()] = - \frac 12 \Ent [\vec x()]
  + \frac 12 \Frn [\vec x()] \label{eq:act-+}
\end{align}
yields
\begin{subequations} \label{eq:act.sym}
\begin{align}
  \Ent [\vec x()] =& \int \dd x_i \circ
  (\dif^{-1} \mob)_{ij} \for_j \ , \label{eq:entropy} \\
  \Frn [\vec x()] =& 2 \log \caN^{\dif}_{\tstep} + \frac 12 \int \dd s
  \left[ \dot{x}_i \dif^{-1}_{ij} \dot{x}_j 
    + \for_i (\mob \dif^{-1} \mob)_{ij} \for_j 
    + 2 \mob_{ij} \del_i \for_j \right] \ , \label{eq:frenesy}
\end{align}
\end{subequations}
where the symbol $\circ$ denotes a Stratonovich product
(which is instead understood in $ \dot{x}_i \dif^{-1}_{ij} \dot{x}_j$). The linear
response \eqref{eq:response} can thus be inspected component-wise
according to the decomposition \eqref{eq:act-+}.

\subsection{Time-antisymmetric sector of the path action} \label{sec:entropy}

Note that when the reservoir is in equilibrium, that is $(\dif^{-1}
\mob)_{ij} = \delta_{ij}/T$, the time-antisymmetric term
\eqref{eq:entropy} becomes $\Ent [\vec x()] = (1/T) \int \dd x_i \circ
\for_i$. According to stochastic energetics, the work done by the
reservoir force ($-\for_i$) is equal to the heat $Q = -\int \dd x_i
\circ \for_i$ absorbed by the system, thereby rendering $\Ent [\vec
  x()]$ as the entropy change $-Q/T$ of the reservoir by Clausius'
equality. If the inverse temperature matrix were not diagonal, but
rather had a diagonal form $(\dif^{-1} \mob)_{ij} = \delta_{ij}/T_{i}$
(no summation) as, \eg., when there is no hydrodynamic correlation
between the degrees of freedom and each one is thermalized by a
reservoir that is in equilibrium at a different temperature $T_i$,
then the time-antisymmetric part of the action is found similarly as
$-\sum_i Q_i/T_i$: Thanks to each reservoir being separately at
thermal equilibrium, the total change of entropy in all reservoirs
would still be given by a sum of Clausius equalities.

The situation considered in this paper instead does not fall into the
previous categories.  The reservoir is out of equilibrium in a more
drastic (albeit quantifiable) way than multiple decoupled reservoirs
thermalizing uncorrelated degrees of freedom. This is captured in the
non-diagonality of the inverse temperature matrix $(\dif^{-1}
\mob)_{ij}$. Under these circumstances, the time-antisymmetric part
\eqref{eq:entropy} of the action does not consist of heat exchanges
occurring at any well-defined temperature, and thus equilibrium
thermodynamic relations become of little utility in identifying
\eqref{eq:entropy} with the entropy produced in the reservoir.  In the
particular case of our experimental paradigm where the inverse
temperature matrix is given in \Eqref{eq:itm}, the entropic part of
the action \eqref{eq:entropy} consists of three terms,
\begin{align} 
\Ent [\vec x()] =&
(\dif^{-1}\mob)_{11} \int \dd x_1 \circ \for_1
+ (\dif^{-1} \mob)_{22} \int \dd x_2 \circ \for_2 
+ (\dif^{-1} \mob)_{12} \int \dd x_1 \circ \for_2.
\end{align}
The first two terms, which equal $-Q_1/(T+T_{\rm em}) - Q_2/T$, seem
to suggest that the ``warmer'' particle (subject to artificial noise)
exchanges heat at the temperature $T+T_{\rm em}$ with the reservoir,
while the other does so at the cooler solvent temperature $T$, but
this simple identification of the warmer and cooler temperatures would
break down if more than one particle were subject to the artificial
noise.\footnote{If the two particles were subject to different nonzero
  noise amplitudes---diagonal entries $\gamma T_{\rm em}^{(i)}$ in
  \Eqref{eq:Dem}---the inverse temperature matrix would be found as \[
  \Dif^{-1} \Mob = \frac {1} {T^2 + TT_{\rm em}^{(1)} +TT_{\rm em}^{(2)}
    + (1-\epsilon^2)T_{\rm em}^{(1)}T_{\rm em}^{(2)} } \begin{bmatrix}
    T+T_{\rm em}^{(2)} & -\epsilon T_{\rm em}^{(1)} \\ -\epsilon
    T_{\rm em}^{(2)} & T + T_{\rm em}^{(1)}\end{bmatrix} \ .\]}

More importantly, however, we have also the cross-term $\int \dd x_1
\circ \for_2$ scaled by a temperature-like quantity $-T (T+T_{\rm em})
/ \gamma \mob_{12} T_{\rm em}$.  One observes that its inverse becomes
zero when there is no external noise ($T_{\rm em} \mathbin{=}0$) or no
hydrodynamic coupling ($\mob_{i\neq j} \mathbin{=}0$).  That is, the
presence of this strange cross-term owes to the combination of
hydrodynamic coupling and external noise, and not a single one of
these effects on its own. As such, the product $\dot{x}_i \for_j$ with
$i \mathbin{\neq} j$ somewhat quantifies this particular manner of
departure from equilibrium, whereas it does not bear a clear
mechanical meaning, for instance as the power delivered by degree of
freedom $j$ on $i$. The latter interpretation was adopted in
Ref.~\cite{ber14}, in particular as a heat flux between particles, but
we should note that Appendix 4.7.2 of Ref.\ \cite{sek10} suggests a
definition of heat in hydrodynamically coupled systems that does not
match that used in Ref.~\cite{ber14}.

We are finally left with the issue of consistently referring to $\Ent
[\vec x()]$. Having argued above in what ways it is \emph{not}
identifiable with an equilibrium thermodynamic entropy production, one
should note that most notions of entropy extended into the
nonequilibrium regime are bound to detach from
equilibrium-thermodynamical underpinnings to some extent~\cite{mae12}.
This encourages us to stick with the word ``entropy'' referring to
time-antisymmetric part of the action, $\Ent [\vec x()]$, alluding to
its significance in the sense of a distinguishability between two
opposite flows of time.

\subsection{Time-symmetric sector of the path action}

In this case we can be more liberal with names, as the time-symmetric
component has no classic counterpart in equilibrium statistical
mechanics.  The term \emph{frenetic} and the name \emph{frenesy} have
been recently proposed to describe the time-symmetric part of the
action because of its relation to the dynamics' volatility against
remaining in a given neighborhood in state space~\cite{bai10,bas15}.
Due to its relationship with the mean number of jumps expected in a
system subject to jump dynamics, it is a sort of \emph{mean dynamical
  activity}.  For the linear response (\ref{eq:response}) applied to
{\em equilibrium} conditions, it can be shown
\cite{bai09,bai09b,bai13} that the excess in this time-symmetric
property and the excess in the entropy production make equal
contributions, giving rise to the celebrated Kubo formula \cite{kub66}
wherein the response to a perturbation is characterized
\emph{entirely} by the heat dissipated into the environment in the
course of accommodating for the perturbing force. For a system out of
equilibrium, however, the response arising from the the symmetric and
antisymmetric sectors differ from each other. Thus, the frenetic and
entropic parts in general manifest distinctly different statistical
properties of the microscopic dynamics underlying the system.

\section{Linear response: deterministic perturbation} \label{sec:resp.f}

In this section we specialize the scheme outlined in the previous
section to derive the linear response to a small variation, $\for_i
\rightarrow \for_i + \dfor_i$, of the deterministic forces acting on
the system. In previous works~\cite{bai09, bai09b} one may find a
scheme suitable for nonequilibrium systems with hydrodynamic
interactions.  Nevertheless, it was assumed that the environment
satisfies the second fluctuation theorem $\dif_{ij} = T \mob_{ij}$,
namely that it is an equilibrium heat bath.  Here, the external noise
makes this invalid, and we modify those results to accommodate the
particular physical scenario discussed in this article.

\subsection{Response formulas}

We have already written the action---\Eqref{eq:action} or
\eqref{eq:act.sym}---in the Stratonovich convention for
state-independent coefficients. The first variation of this action
with respect to a variation of the force is found simply as the first
order term in $\dfor_i$ after replacing $\for_i\rightarrow \for_i +
\dfor_i$:
\begin{subequations} \label{eq:act.sym.f}
\begin{align}
  \delta \Ent [\vec x()] =& \int \dd x_i \circ
  (\dif^{-1} \mob)_{ij} \dfor_j \ , \label{eq:entropy.f} \\
  \delta \Frn [\vec x()] =&  \int \dd s
  \left[ \for_i (\mob \dif^{-1} \mob)_{ij} \dfor_j 
    + \mob_{ij} \del_i \dfor_j \right] \ . \label{eq:frenesy.f}
\end{align}
\end{subequations}
Earlier, we have alluded to the fact that without a diagonal inverse
temperature matrix $(\dif^{-1} \mob)_{ij}$, the entropic term ($-$)
does not admit an equilibrium-thermodynamic interpretation in terms of
temperature-scaled (excess) heat exchanges with the
environment. Similarly, the frenetic term ($+$) does not admit an
obvious dynamical interpretation the way it does in the case of a
trivial inverse temperature matrix (proportional to identity) and
conservative perturbation \cite{bai09b}. It is possible to see where
the difficulty lies, by rewriting the integrand of
\Eqref{eq:frenesy.f} in the form $\left[ (\mob \for)_j + \dif_{ij}
  \del_i \right] (\dif^{-1} \mob \dfor)_j$. Only when $(\dif^{-1} \mob
\dfor)_j$ is equal to the gradient $\del_j \psi (\vec x)$ of a
function, the integrand could be written as $\genL \psi (\vec x)$
(with $\genL = (\mob \for)_j \del_j + \dif_{ij} \del_i \del_j$ being
the backward generator of the dynamics), which is a measure of the
instantaneous propensity of the system towards different values of the
``potential'' $\psi( \vec x(s))$. However, the condition of
$(\dif^{-1} \mob)_{ij} \dfor_j$ being a gradient has no general
validity.

If furthermore the variation $\dfor_i (\vec x (s) , \lambda (s))$ is
expressed in terms of an external parameter $\lambda (s)$, then
connection to a response function $\rf_{\caO}^{\lambda} (t,s)$ can be
made through the relation $\dfor_i = \delta \lambda (\del \for_i
/\del \lambda)$.\footnote{This is true if $\dfor_i(\vec x, \lambda)$
  does not depend on time derivatives of $\lambda$. We assume this to
  be the case for the sake of simplicity, even though this is not
  essential.}  This yields
\begin{align}
  \rf_{\caO}^{\lambda} (t,s) =& 
  \frac 12 \mean { \caO (t) \left[ \dot{x}_i (\dif^{-1} \mob)_{ij}
    \frac {\del \for_j} {\del \lambda} \right]\!(s) } \nonumber \\
  -&\frac 12 \mean {\caO (t) \left[ \for_i (\mob \dif^{-1} \mob)_{ij}
      \frac {\del \for_j} {\del \lambda}
      + \mob_{ij} \del_i \frac {\del \for_j} {\del \lambda} 
  \right]\!(s)} \ , \label{eq:rf.f}
\end{align}
where the dependence on the earlier time $s$ was denoted in an overall
fashion to indicate that everything inside the preceding square
brackets is to be evaluated at that time. The first and second lines
are the entropic and frenetic parts of the response, respectively.

\subsection{Numerical experiments}

We performed numerical simulations of the experimental setup described
in Sec.\ \ref{sec:expara} by generating many random trajectories
according to its overdamped Langevin dynamics description given
therein. Susceptibilities can be measured directly by running dynamics
perturbed by a small constant parameter as well as unperturbed
dynamics, and taking the difference in an observable of interest. The
susceptibilities $\sus (t) = \int_0^t \dd s\, \rf (t,s)$ predicted by
the linear response theory are instead calculated only over the
unperturbed trajectories. In principle, this scheme can be employed in
actual experiments which are able to record the stochastic
trajectories of the particles with reasonable time-resolution.

\subsubsection{Perturbing the cooler trap}

In our first example, we have the unperturbed force $\for_i
\mathbin{=} - \kappa_{i} x_i$ and we consider a perturbation $\dfor_i
\mathbin{=} - \delta \kappa_{i} x_i$ on top of it. For the sake of
having a scalar response function, let us assume $\delta \kappa_1
\mathbin{=} 0$ and $\delta \lambda \mathbin{=} \delta
\kappa_{2}$. That is, the ``cooler'' trap without the artificial noise
is being perturbed. We then have $\del \for_j / \del \lambda
\mathbin{=} -\delta_{j2} \delta x_2$ for use in (the integral of)
\Eqref{eq:rf.f}. As the observable of interest, we focus on the energy
stored in the traps; the total energy $U (\vec x) = (1/2) \kappa_i
x_i^2$ and the energy $U_1 (x_1) = (1/2) \kappa_1 x_1^2$ stored in the
first trap. In other words, we observed the susceptibilities
\begin{align}
  \sus_{U}^{\kappa_{2}} (t) = \frac {\del} 
  {\del \kappa_{2}} \mean{U(\vec x(t))}\quad \text{and} \quad
  \sus_{U_1}^{\kappa_{2}} (t) = \frac {\del }
  {\delta \kappa_{2}} \mean{U_1( x_1(t))} \ ,
\end{align}
where the latter is interesting in that the first trap feels the
perturbation on the second only via hydrodynamic coupling. The
response in the first trap can be found simply by subtracting,
$\sus_{U}^{\kappa_2} - \sus_{U_1}^{\kappa_2}$.

\begin{figure} \centering
  \begin{tabular}{r}
    \includegraphics[scale=1]{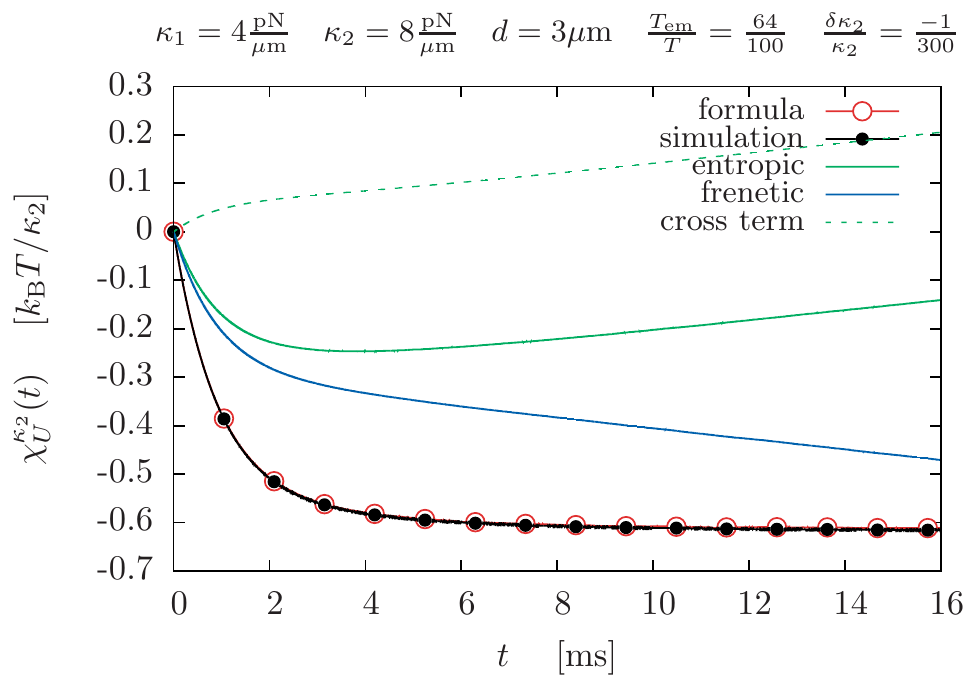}\\
    \includegraphics[scale=1]{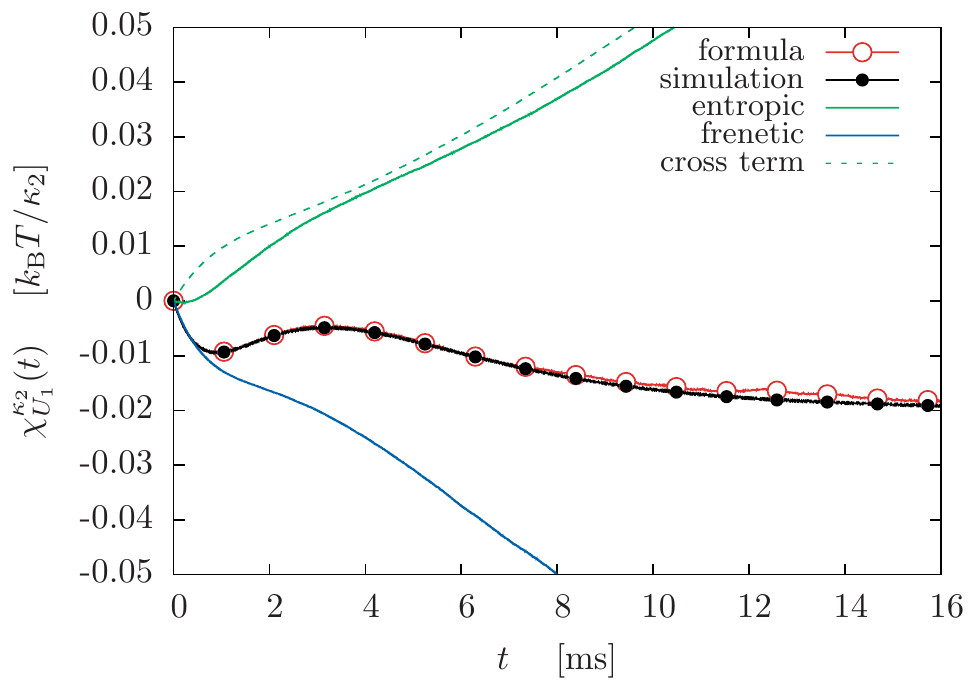}
  \end{tabular}
  \caption{Response formula compared to results of simulation with
    Heun integration ($\tstep=\gamma/4000\kappa_{1}$). (top)
    Susceptibility $\sus_{U}^{\kappa_{2}} (t)$ of total energy to a
    stepwise perturbation $\delta \kappa_{2}$ of the stiffness of the
    second trap. (bottom) Susceptibility $\sus_{U_1}^{\kappa_{2}} (t)$
    of the energy in the unperturbed trap. Simulation parameters were
    given at the top, in units appropriate for the experimental
    situation. The averages were calculated from about 50 million
    trajectories. The green and blue graphs represent the entropic and
    frenetic parts adding up to the total response in red. The dashed
    green graph is the specific contribution of the cross term $\dot{x}_1
    \dfor_2$.} \label{fig:U.k2}
\end{figure} 

Graphs for the susceptibilities are shown in Fig.~\ref{fig:U.k2},
along with some technical details. At the initial time $t \mathbin{=}
0$ when the perturbation was turned on, the positions were effectively
sampled from the nonequilibrium stationary state created by the
presence of external noise on the first trap. In the example shown,
the second trap was softened ($\delta \kappa_2 < 0$), resulting in
more spread of the particles, hence a larger positive value for the
energies, and therefore a negative susceptibility after division by
$\delta \kappa_2$. In the bottom graph, we see how the first
particle's energy responds to a perturbation in the other trap.  In
particular, softening the cooler trap is seen to add energy to the
warmer trap, via hydrodynamic coupling. The entropic and frenetic
parts stray from each other in both observables.  We have also plotted
(dashed line) the part of the entropic response that is proportional
to the off-diagonal entry $(\dif^{-1} \mob)_{12}$ (alternatively,
according to \Eqref{eq:entropy.f} or \eqref{eq:rf.f}, this part arises
from $\dot{x}_1 \dfor_2$). This cross-term appears to account for most
of the response in the unperturbed trap, which is intuitive since the
fluctuations here have to do eventually with the cross-correlation
between the traps. In addition, the overall running-away between the
entropic and frenetic parts appear to also be induced by this cross
term.  This point is further discussed in the following complementary
case.

\subsubsection{Perturbing the warmer trap}

In the above example, we considered the case of the strength of the
cooler optical trap being manipulated, where the perturbing force had
$\dfor_1 \mathbin{=} 0$ and $\dfor_2 \mathbin{\neq} 0$. Note that the
nonzero component $\dfor_2$ multiplies the off-diagonal entry
$(\dif^{-1} \mob)_{12}$ of the inverse temperature matrix in products
of the form $(\dif^{-1} \mob)_{ij} \dfor_j$ of
\Eqref{eq:act.sym.f}. As this off-diagonal entry is the main feature
that distinguishes this nonequilibrium situation from those with
decoupled reservoirs, we chose to deal with the case where this term
is present first. However, the opposite case of manipulating the
warmer trap has $\dfor_2 \mathbin{=} 0$, which removes the
off-diagonal entry $(\dif^{-1} \mob)_{12}$ from the evaluation of the
response. Simulations pertaining to this case
illustrate the distinction.

\begin{figure} \centering
    \includegraphics[scale=1]{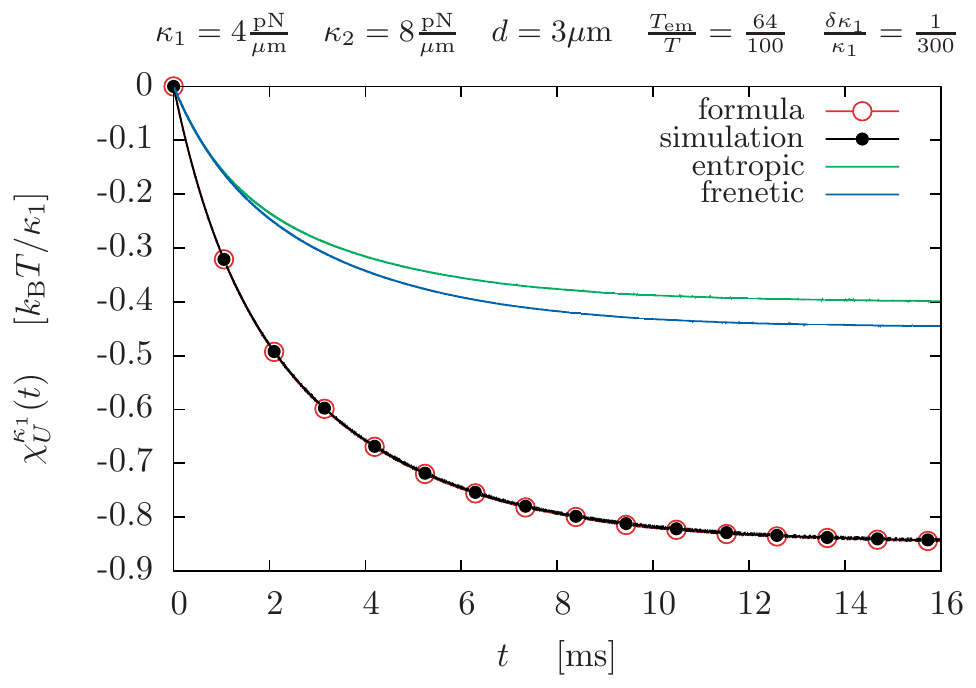}
    \\ \includegraphics[scale=1]{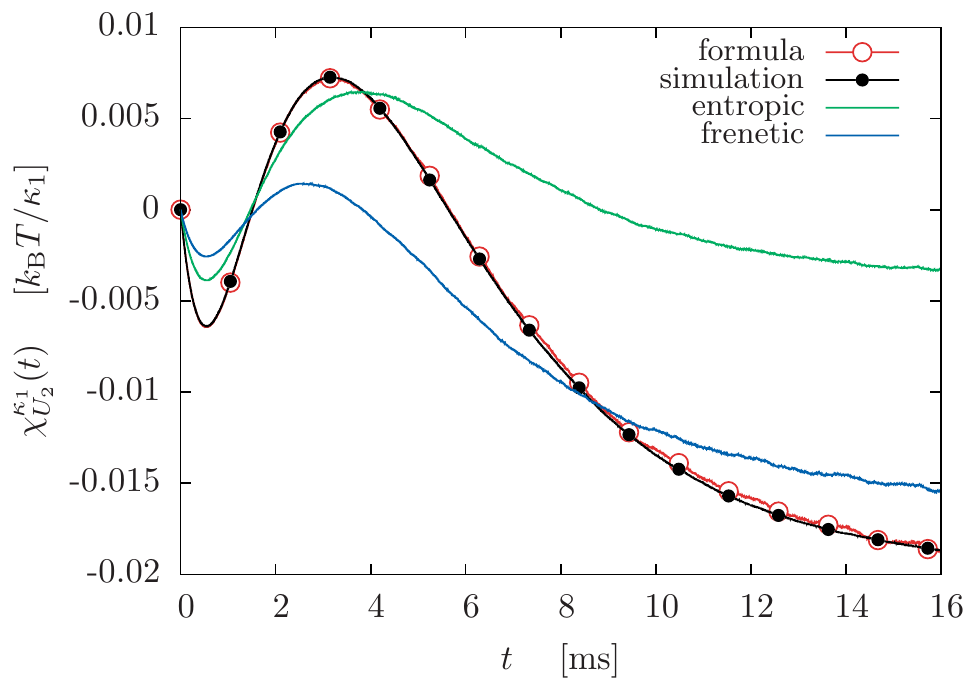}
  \caption{Response formula compared to results of simulation with
    Heun integration ($\tstep=\gamma/4000\kappa_{1}$). (top)
    Susceptibility $\sus_{U}^{\kappa_{1}} (t)$ of total energy to a
    stepwise perturbation $\delta \kappa_{1}$ of the stiffness of the
    second trap. (bottom) Susceptibility $\sus_{U_2}^{\kappa_{1}} (t)$
    of the energy in the unperturbed trap. Simulation parameters were
    given at the top, in units appropriate for the experimental
    situation. The averages were calculated from about 50 million
    trajectories. The green and blue graphs represent the entropic and
    frenetic parts adding up to the total response in red, which in
    this case follow similar trends as opposed to that depicted in
    Fig.~\ref{fig:U.k2}.}
   \label{fig:U.k1}
\end{figure} 

Fig.~\ref{fig:U.k1} depicts the associated susceptibilities of the
total energy and the energy in the unperturbed trap, similar to the
case above. Due to the softness of the warmer trap---which was an
arbitrary choice of ours---the relaxation is seen to be slower than in
the previous case. Here, we note that the runaway behaviour seen in
the previous Fig.~\ref{fig:U.k2}, of both the entropic and the
frenetic parts, disappears along with the disappearance of the cross
term $(\dif^{-1} \mob)_{12}$ in the inverse temperature matrix. For
the entropic term, this corresponds to the removal of the cross term
$\dot{x}_1 \dfor_2$ from the entropy production, the correlations with
which had been depicted in dashed lines previously.

Following the discussion in Sec.~\ref{sec:entropy},
it is possible to argue that, at least for the case of separable
potential $U(\vec x) = \sum_i U_i (x_i)$ that we have, this runaway
behaviour, when it is present, has to do with the fact that the cross
terms $\dot{x}_i \for_j$ ($i \mathbin{\neq} j$) constitute the
residual housekeeping entropy production rate associated with the
stationary nonequilibrium imposed by the broken proportionality
between $\dif_{ij}$ and $\mob_{ij}$: Each term of $\dot{x}_i \for_i$
equals the rate of change $-\dot{U}_i (x_i)$ of the energy stored in
trap $i$, which vanishes on the average in stationarity by definition,
thereby rendering the remainder $\dot{x}_i \for_j$
($i \mathbin{\neq} j$) proportional to the housekeeping dissipation on
the average. In the present case of perturbing the warmer trap, it is
the absence of correlations with (the excess in) this residual steady
rate of entropy production that is apparently responsible for the
disappearance of the runaway behaviour that was seen in the previous
example.  Similarly, the frenetic part is also stabilized or not 
depending on the absence of this cross term. 
A notion of frenetic housekeeping cost might be needed to
account for this aspect.

\section{Linear response: thermal perturbation}
\label{sec:resp.T}

Here, we discuss the linear response to small variations in
temperature, or equivalently the diffusion coefficients, in a scheme
where we discretize time. A similar scheme was used in Ref.\ \cite{bai14}
for a single degree of freedom and adopting the It\^o convention
rather than the Stratonovich one. In the Appendix we discuss the unusual
issue of needing to select from the start which convention to use, 
even if the stochastic equations have a constant noise prefactor.

We have seen above that how the continuum limit $\tstep \rightarrow 0$
is taken becomes irrelevant as far as the response to mechanical
perturbations was concerned, since the force-dependent terms of the
action \eqref{eq:action} are well-behaved. In contrast, we will see
that varying the temperature or diffusion coefficients retains
regularization-dependent parts of the action. 

\subsection{Response formulas}

The variation of the action, \Eqref{eq:action} or \eqref{eq:act.sym},
with respect to the diffusion matrix can be done as before by
substitution ($\dif_{ij} \rightarrow \dif_{ij} + \ddif_{ij}$) and
sifting the linear order. Noting the following
relations,
\begin{subequations}
  \begin{align}
    \delta\! \left( \Dif^{-1} \right) =& 
    -\Dif^{-1} \, \dDif \, \Dif^{-1} \ , \label{eq:dDi}\\
    \delta \operatorname{Tr} \log \Dif =& \operatorname{Tr} 
    \!\left[ \Dif^{-1} \, \dDif \right] 
    = - \operatorname{Tr} \!\left[ \delta (\Dif^{-1}) \Dif \right] \ ,
    \label{eq:dlogD}
  \end{align}
\end{subequations}
which can be confirmed easily by differentiating the definition
$\Dif^{-1} \Dif = \mathbf{1}$, and using a Taylor series around
$\mathbf{1}$ for the logarithm, we find
\begin{subequations}
\begin{align}
  \delta \Ent [\vec x()] =& \int \dd x_i \circ 
  (\ddi \mob)_{ij} \for_j \label{eq:entropy.T} \\
  \delta \Frn [\vec x()] =& \sum_{\tstep}
  \left( \frac{\Delta x_i \Delta x_j} {2 \tstep}
  - \dif_{ij} \right) \ddi_{ij}  
  +\frac 12 \int \dd s\, 
  \for_i (\mob \ddi \mob)_{ij}\for_j \ . \label{eq:frenesy.T}
\end{align}
\end{subequations}
Here, we have denoted $\delta (\dif^{-1})_{ij}$ as $\ddi_{ij}$, and
have refrained from substituting \Eqref{eq:dDi}, both for the sake of
compactness. Also note that, similarly to the force perturbation case,
the perturbation $\ddif_{ij}$ is in general a function of
time. Finally, the first term of the symmetric part
\eqref{eq:frenesy.T}---that which depends on the regularization---has
been written as a discrete sum over time slices, \emph{before} the
continuum limit is taken, in the spirit of \Eqref{eq:norm}. In the
limit $\tstep \rightarrow 0$, the term $\Delta x_i \Delta x_j
\rightarrow 2 \dif_{ij} \tstep$ as a result of the equation of motion
\emph{to lowest order} in $\tstep$, thereby negating the singularity
coming from the normalization. Hence, the regularization-dependent
term is expected to be at least as well-behaved as the stochastic
integral of the entropic part \eqref{eq:entropy.T}, with each
increment being of the order $\tstep^{\frac 12}$.

The response function then has the form
\begin{align} 
  \rf_{\caO}^{\lambda} (t,s) =& 
  \frac 12 \mean { \caO (t) \left[ \dot{x}_i 
      \frac {\del \dif^{-1}_{ij}} {\del \lambda} 
      (\mob \for)_j \right]\!(s) } \nonumber \\
  -&\frac 12 \mean {\caO (t) \left[ \dot{\zeta}_{ij} 
      \frac {\del \dif^{-1}_{ij}} {\del \lambda} + \frac 12
      (\mob \for)_i \frac {\del \dif^{-1}_{ij}}
      {\del \lambda} (\mob \for)_j
       \right]\!(s)} \ , \label{eq:rf.T}
\end{align}
where we have used a definition
\begin{align}
  \dot{\zeta}_{ij} = \lim_{\tstep \to 0} \frac{1}{\tstep}
  \left( \frac {\Delta x_i \Delta x_j} {2 \tstep} - \dif_{ij} 
  \right) \ , \label{eq:regdep}
\end{align}
for the regularization-dependent part. For the scenario at hand, 
a natural scalar parameter $\lambda (t)$ to perturb is the
effective temperature $T_{\rm em}$.

\subsection{Numerical experiments}

In order to check the response formulas above, we performed numerical
simulations considering a perturbation
$T_{\rm em} \to T_{\rm em} + \delta T_{\rm em}$ applied in a stepwise
fashion on our paradigmatic system in a stationary state corresponding
to the original effective temperature $T_{\rm em}$. Considering the
total energy $U(\vec x) = (1/2) \kappa_{i} x_i^2$ as the observable,
one obtains a susceptibility akin to a heat capacity by integrating
the response function \eqref{eq:rf.T}. The inverse of the diffusion
matrix \eqref{eq:Dif.2} for our example is found as
\begin{align}
  \Dif^{-1} = \frac 1T \Mob^{-1} - \frac \gamma T 
  \begin{bmatrix}  \frac {T_{\rm em}} {T+T_{\rm em}} &0\\0&0
    \end{bmatrix} \ ,
\end{align}
which implies (with $\lambda = T_{\rm em}$)
\begin{align}
  \frac {\del \dif^{-1}_{ij}} {\del \lambda} = 
  \begin{bmatrix} -\frac \gamma {(T+T_{\rm em})^2} &0\\0&0
    \end{bmatrix} \ ,
\end{align}
to be used in \Eqref{eq:rf.T}.  

\begin{figure} \centering
    \includegraphics[scale=1]{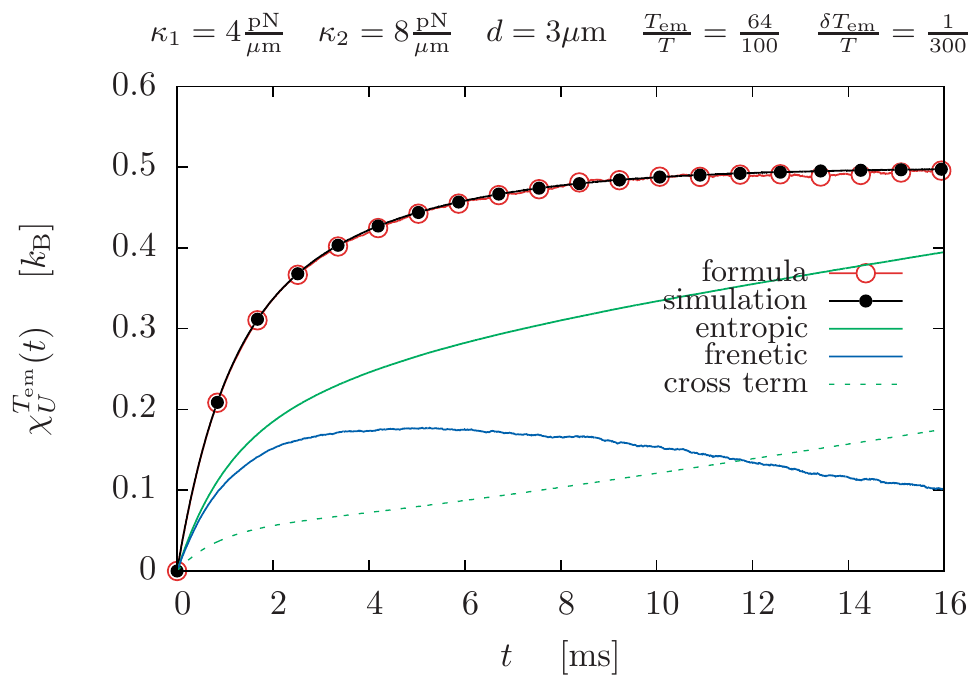}
    \includegraphics[scale=1]{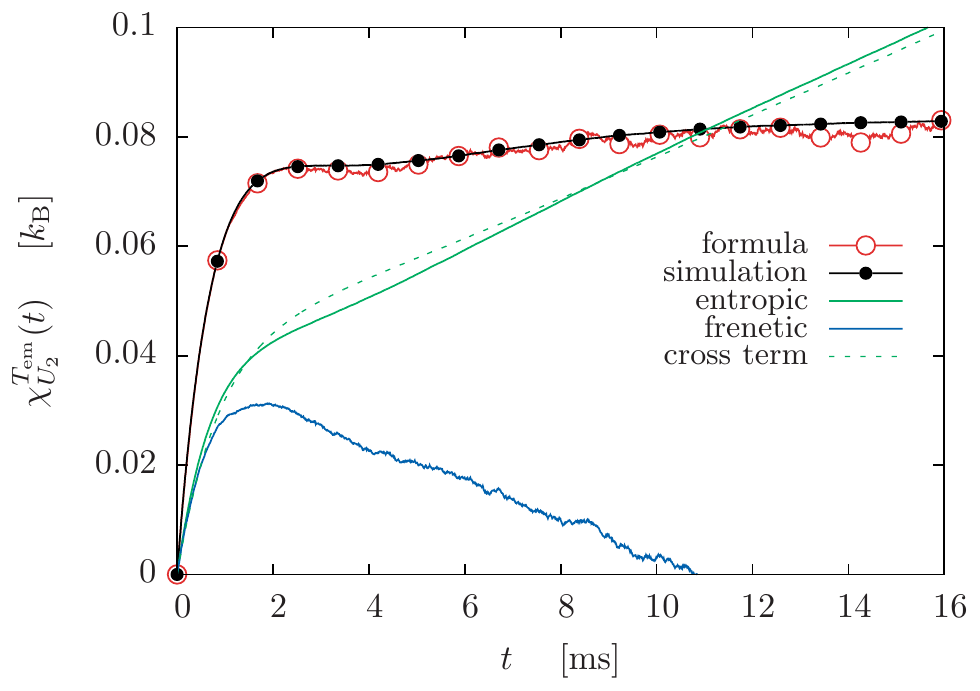}
    \caption{Response formula compared to results of simulation with
      Heun integration ($\tstep=\gamma/500\kappa_{1}$). Depicted are
      (top) the ``heat capacities'' for the total energy
      $\sus_{U}^{T_{\rm em}}$, and (bottom) the energy in the
      unperturbed trap $\sus_{U_2}^{T_{\rm em}}$. Sampling was done
      over 100 million trajectories. Simulation parameters were
      given at the top, in units appropriate for the experimental
      situation. The green and blue graphs represent the entropic and
      frenetic parts adding up to the total response in red. The
      dashed green graph is the contribution of the cross-term
      $\dot{x}_1 \for_2$.} \label{fig:U.T}
\end{figure}

In the simulations, as before, we measured the susceptibilities of
both the total energy $U(\vec x) = (1/2) \kappa_{i} x_i^2$, and the
energy $U_2 (x_2) = (1/2) \kappa_{2} x_2^2$ in the unperturbed trap,
to validate the response formulas. In other words, we verified the
susceptibilities $\sus_{U}^{T_{\rm em}} (t)$ and $\sus_{U_2}^{T_{\rm
    em}} (t)$.  The second susceptibility, as before, is interesting
in that it is only due to hydrodynamic coupling that there is a
nonzero response in this trap, as there is no mechanical coupling
between the two particles. In Fig.~\ref{fig:U.T} we see the
susceptibilities upon an increase in the effective temperature $T_{\rm
  em}$ of the external noise. The increase in the energy stored in the
cooler trap due to the increased temperature difference occurs faster,
since the cooler trap was arbitrarily chosen to be stiffer.  We
observe again that the entropic and frenetic parts diverge from each
other. The parts of entropic response involving the cross-term
$\dot{x}_1 \for_2$ were plotted in dashed lines, and it is seen, as
before, that this part constitutes most of the response in the
unperturbed trap and also that it is responsible for the run-away of
the entropic part.

\section{Conclusions}

To sum up, we have investigated a physical situation where microscopic
colloidal particles in solvent are held close to each other, enough to
experience hydrodynamically mediated correlations in their
fluctuation. We derived the response to small perturbations of
deterministic or thermal origin, within a path-weight framework.  An
experiment \cite{ber14} that traps the colloids at a more or less
fixed distance is suitable for studying the influence of hydrodynamic
coupling in isolation from other potential complications.  Despite
this simplification, we deal with a system driven out of equilibrium
by unbalanced thermal reservoir influences, in the form of one degree
of freedom being ``heated up'' by an artificial noise.  This is
absorbed into the diffusion matrix of a stochastic equation and leads
to a non-diagonal ``inverse temperature matrix'' $(\dif^{-1}
\mob)_{ij}$.  Hence, the reservoir cannot be thought of as comprising
components that thermalize each degree of freedom separately. A
complex condition thus arises from mixing hydrodynamic interactions
with the action of a nonequilibrium heat bath.  There is no
well-defined equilibrium temperature to scale the heat exchanges that
would yield the change of entropy in the reservoir. Hence, notions as
heat, power delivered, entropy, etc.~need to be considered with care.

We found it more convenient to continue using the term ``entropy
production'' for the time-antisymmetric dimensionless objects emerging
from path-weights. However, this should be considered more a
concession to acknowledging the property of this objects to detect
time-reversal breaking rather than being true equilibrium
thermodynamic entropy changes in the reservoirs.  In particular,
attempting to identify entropy fluxes into reservoirs is compromised
by off-diagonal entries of the inverse temperature matrix endowed by
the nonequilibrium heat bath with hydrodynamic coupling: Cross terms
of the form $\dot{x}_i \for_{j}$ ($i \mathbin{\neq} j$) contribute as
well to our results.  While it is tempting to ascribe a meaning to
such forms along the lines of power transfer between $i$ and $j$, we
have argued that they are rather indicators of the departure from
equilibrium conditions.  Since stationary averages of $\dot{x}_i
\for_i$ do vanish, the non-zero cross term $\mean{\dot{x}_1 \for_2}$
should be in close relationship with the notion of housekeeping
dissipation, necessary to maintain the nonequilibrium condition
$\dif_{ij} \neq T \mob_{ij}$.  It is thus possible to argue that the
measurements of Ref.\ \cite{ber14} can be interpreted as indicators of
the housekeeping dissipation.  Since we deal also with time-symmetric
quantities, possibly concepts as {\em housekeeping frenesy} should be
introduced to pair with that of housekeeping heat.

Within the path probability framework, we derived explicit expressions
for the linear response of such a system to changes in the
deterministic forces as well as thermal influences acting on the
particles. It should be possible to verify these expressions in experiments
measuring particle trajectories \cite{ber14}, just by following the scheme of
our simulations, which perturbed either  the trap strengths
or the magnitude of the artificial noise (temperature).

The fixed distance between the traps allowed an approximation where
diffusion and mobility coefficients were independent of
coordinates. In more general situations, such as the motion of a
collection of active Brownian particles, or colloidal model systems
where particles are not necessarily fixed, this needs to be
incorporated. It is not clear how the phase behaviour or linear
response would be influenced in systems where the coordinate
dependence of the reservoir forces becomes relevant. 

Regarding the regularization-dependence of the temperature response
formulas, Eqs.\ \eqref{eq:rf.T} and \eqref{eq:regdep}, we note that a
reparametrization of the degrees of freedom such as that suggested in
Refs.~\cite{fal16a, fal16b} can remove it. This will be investigated
in a future work.

\ack We thank S. Ciliberto, G. Falasco, and C. Maes for helpful discussions.

\appendix

\section*{Appendix: It\^o and Euler vs. Stratonovich and Heun}

In numerical computations, it is important to ensure consistency
between the numerical integration that generates trajectories and the
interpretation of the stochastic equation of motion \cite{gardiner},
which in turn determines the form of the path weights. Accordingly,
overdamped motion was generated in the simulations of
Ref.~\cite{bai14} via (explicit) Euler integration, matching their
choice of It\^o convention for expressing the path weights. In the
present article, instead, we adopt the Stratonovich convention for
reasons explained at the beginning of
Sec.\ \ref{sec:paction}. Therefore, the numerical integration we used
for generating stochastic trajectories was based on Heun's method,
which is a type of semi-implicit Euler scheme agreeing with the
Stratonovich interpretation, to order $\tstep$
\cite{gardiner,sek10}. However, it seems odd that in a scenario with
  state-independent coefficients this issue would become relevant.

Nonetheless, one can easily check numerically that if an Euler scheme
rather than Heun is used to generate the trajectories, data depart
from our analytical results in the case of temperature
response. Furthermore, and somewhat surprisingly, it is not possible
to shift between the two noise interpretations in the final
expressions for the response in order to recover agreement: Notice
that due to its lack of temperature dependence, the temperature
derivative erases the last term of the path action \eqref{eq:action},
which however has to be involved in any alteration of the noise
interpretation. Even though a conversion of the stochastic integrals
is still permissible after this point, the trajectories must be
generated by numerics respecting the original convention. Indeed, the
overdamped trajectories of Ref.~\cite{bai14} were generated using an
Euler scheme notwithstanding the final conversion of stochastic
integrals from It\^o to Stratonovich, which was done in order to
define the time-symmetry of the integrals.

A heuristic justification of such discrepancy despite
  the state-independence of noise coefficients is the following:
Typically, agreement between the noise interpretation and stochastic
integration is sought in the path increments $\Delta x_i$ to order
$\tstep$.  However, in our scheme we deal with path
weights that eventually rely on the \emph{square} of path increments
to order $\tstep^2$. As a result, any disagreement between the
stochastic interpretation and integration at order $\tstep^{3/2}$ of
the path increment $\Delta x_i$ will introduce error, even if constant
coefficients ensure agreement at order $\tstep$. Although we have not
carried out a rigorous proof showing what the terms of order
$\tstep^{3/2}$ might be, it appears that the same rule of thumb
prescribing semi-implicit integration to accompany Stratonovich
interpretation appears to remain valid.

\bibliographystyle{iopart-num}
\section*{References}
\bibliography{bib_noneq_150808}

\end{document}